\documentclass[journal]{igtesymp}
\ifCLASSINFOpdf
\else
\fi
\usepackage[tight,footnotesize]{subfigure}

\usepackage{printlen}

\usepackage{hyperref}
\usepackage{graphicx}
\usepackage{amsmath}
 \usepackage{tikz}
 \usetikzlibrary{arrows.meta}

\usepackage{lipsum}
\usepackage{printlen}

\usepackage{color}

\usepackage{pgfplots}
\pgfplotsset{compat=1.18}
\usetikzlibrary{arrows.meta}
\usepackage{booktabs}
\usepackage{siunitx}

\usepackage{bm}

\newcommand{\av}[1]{\bm{\MakeLowercase{#1}}}
\newcommand{\mtr}[1]{\bm{#1}}

\newcommand{\Material}[1]{\underline{\bm{\MakeLowercase{#1}}}}
\newcommand{\tp}[1]{#1^\top}

\makeatletter
\def\ps@igtedisclaimer{%
  \def\@oddhead{}\def\@evenhead{}%
  \def\@oddfoot{\hss\normalfont\footnotesize\begin{tabular}[t]{@{}c@{}}%
    This work has been submitted to the 22nd International IGTE Symposium 2026 for possible publication.\\
    Copyright may be transferred without notice, after which this version may no longer be accessible.%
  \end{tabular}\hss}%
  \let\@evenfoot\@oddfoot}
\makeatother

\begin{document}
%
\title{Energy-Based Anisotropic Hysteresis Modeling for Permanent Magnets}
%
%
%
\author{Walter~Friza\IEEEauthorrefmark{1},  Manfred~Kaltenbacher\IEEEauthorrefmark{1}
 \\ \vspace{0.3cm} \normalsize{
\IEEEauthorblockA{\IEEEauthorrefmark{1}Institute for Fundamentals and Theory in Electrical Engineering, Inffeldgasse 18/1, A-8010 Graz\\
 E-mail: walter.friza@tugraz.at}}
\thanks{Manuscript received November, 2026; Corresponding author: Walter Friza (email: walter.friza@tugraz.at).}%
}

%
%


\IEEEaftertitletext{\vspace{-1cm}\noindent\begin{abstract} Modeling the anisotropic hysteresis behaviour of permanent magnets requires vector hysteresis models capable of representing preferred (easy) magnetization directions. This work extends the energy-based vector play model to anisotropic materials by replacing the spherical pinning region of each particle with an ellipsoid, whose semi-axes represent direction-dependent pinning forces.
Special attention is paid to the geometrically consistent adaptation of the update formula for the vector play model.
The anhysteretic part of the model is also given an angular dependency. 
The model is compared to measurements taken from an NdFeB magnet and reproduces the measured loops for field inclinations of up to 30° from the easy axis.
\end{abstract}
\noindent\vspace{\baselineskip}}

\maketitle
\thispagestyle{igtedisclaimer}\pagestyle{empty}

%
\IEEEpeerreviewmaketitle

\section{Introduction}
Permanent magnets (PMs) are essential components of electrical machines, actuators, and sensors. Designing these devices requires finite element (FE) simulations to predict operating points and irreversible demagnetization under overload, short-circuit, or magnetizing conditions.
Commercial FE tools typically model PMs using the linear relationship $\av{b} = \av{b}_\mathrm{r} + \mu_0\mu_\mathrm{r}\av{h}$ and nonlinear demagnetization models have been developed on this basis~\cite{Ruoho2007, Moree2022}. 
Most of these nonlinear models, however, are scalar along the easy axis. 
The transverse direction is either treated as linear or via lookup tables, and recoil follows ad hoc rules. Even the most comprehensive anisotropic nonlinear models developed for AlNiCo are formulated in two dimensions (2D)~\cite{Ibrayeva2023, Westerberg2024}.

Since most PM materials are highly anisotropic, neglecting the full vector character of the material response is a serious limitation. Due to their uniaxial texture, the remanence, coercivity, and permeability differ significantly between the easy axis and the transverse plane. The switching field also depends on the angle between the applied field and the easy axis. 
The angular dependence of the switching field varies depending on the material and microstructure. It ranges from pinning-dominated switching to coherent rotation, which is described by the Stoner--Wohlfarth model~\cite{Moree2022, Moree2023Energies}. 
Fields inclined to the easy axis are particularly harmful because the higher transverse permeability provides a flux path for the external field. In addition, magnetization along the transverse direction reduces the easy-axis magnetization~\cite{Westerberg2024}.

Capturing these effects consistently requires a 3D vector description. In real devices, the field in a magnet is rarely confined to a plane. End regions, skewed machines, axial-flux machines, Halbach arrays, and magnetizing fixtures all produce field components in all spatial directions. 

Although vector hysteresis models exist, their application to PMs is limited. Both Preisach-type and Jiles--Atherton-based approaches have been applied to PMs, but they are largely scalar or empirical~\cite{Moree2023Energies}. Vector Preisach-type models have mainly been developed for soft magnetic materials~\cite{Mayergoyz2003}. These approaches are often computationally expensive.

The energy-based (EB) hysteresis model is an attractive alternative~\cite{bergqvist1997magnetic, Henrotte2006, FrancoisLavet2013}. 
Derived from an energy balance, the EB model is thermodynamically consistent and vectorial by construction. Its constitutive law leads to well-posed FE formulations that can be solved efficiently~\cite{Domenig2024FEformulation}.
Although anisotropic extensions of the EB model exist~\cite{Steentjes2017,Upadhaya2020}, they have only been developed and validated for electrical steel sheets. To the best of the authors’ knowledge, no anisotropic EB model has been reported for hard magnetic materials.

This work therefore presents a three dimensional (3D) anisotropic EB material model for permanent magnets. Uniaxial anisotropy and angle-dependent switching are incorporated via the anhysteresis and geometrically consistent pinning description.

The rest of the article is structured as follows: First, the energy-based hysteresis model is introduced. Next, the anisotropic extension of the vector play model is presented. Finally, the model is tested against measurements.

\section{Energy-Based Hysteresis Model}\label{sec:EB_model}
The energy-based modeling approach builds upon the laws of thermodynamics and yields the constitutive law
\begin{equation}
    \av{B} = \mu_0\av{H} + \av{j}~,
\end{equation}
with $\av{b}$ the flux density, $\av{H}$ the field strength and $\av{j}$ the polarization.
It uses a dry-friction analogy~\cite{bergqvist1997magnetic} and splits the applied field into a reversible and an irreversible part,
\begin{equation}
  \av{H} = \av{H}_{\mathrm{rev}} + \av{H}_{\mathrm{irr}}~.
  \label{eq:split}
\end{equation}
The vector play model (VPM) is an approximation of the energy-based model and gives an explicit update rule for the reversible field
\begin{equation}\label{eq:isotropic_vpm_H_rev}
    \av{H}_\mathrm{rev} = 
    \begin{cases} \av{H}_\mathrm{rev,old} & \lVert\av{H}-\av{H}_\mathrm{rev,old}\rVert \leq \kappa \\
    \av{H} - \kappa\frac{(\av{H}-\av{H}_\mathrm{rev,old})}{\lVert\av{H}-\av{H}_\mathrm{rev,old}\rVert} & \lVert\av{H}-\av{H}_\mathrm{rev,old}\rVert > \kappa
    \end{cases}
    ~,
\end{equation}
with a pinning force $\kappa$.
The polarization is obtained by evaluating the anhysteretis function $J_\mathrm{anh}$ at the reversible field
\begin{equation}
  \av{j} = J_\mathrm{anh}\bigl(\lVert\av{H}_{\mathrm{rev}}\rVert\bigr)\,
               \frac{\av{H}_{\mathrm{rev}}}{\lVert\av{H}_{\mathrm{rev}}\rVert}~.
  \label{eq:anhyst}
\end{equation}
With one pinning force only the major loop can be represented. In order to model also inner hysteresis loops $N$ pinning particles with different strengths $\kappa_l$ are superposed
\begin{equation}
  \av{j} = \sum_{l=1}^{N} \omega_l\, \av{j}_l~,
  \label{eq:sum}
\end{equation}
with weights $\omega_l$, $\sum_l \omega_l = 1$.
For more details we refer to~\cite{Kaltenbacher2026}.

\section{Material Model}
The model described in \autoref{sec:EB_model} is limited to isotropic materials. Extending it to anisotropy requires adapting two components: the pinning mechanism and the anhysteretic function. This section presents an extension of the VPM for each.
\subsection{Anisotropic Pinning Particles}
\begin{figure}[!ht]
    \centering
\begin{tikzpicture}[>=Latex, line width=1pt]
\draw[fill=blue!20, draw=blue!60!black, thick]
    (0,0) ellipse (2cm and 1cm);

\draw[draw=blue!60!black, thin]
    (0,0) ellipse (0.5cm and 1cm);

\draw[draw=blue!60!black, thin, dashed]
    (0,0) ellipse (2cm and 0.4cm);

\draw[-latex, thick] (-0,0) -- (2,0) node[midway, above, yshift=-0.1em] {$\kappa_x$};

\draw[-latex, thick] (0,0) -- (-0.47,-0.4) node[midway, below, yshift=-0.4em, xshift=-1.1em] {$\kappa_z$}; 

\draw[-latex, thick] (0,0) -- (0,1) node[above] {$\kappa_y$};

\coordinate (O) at (0,0);
\coordinate (P) at (-1.775,0.45);
\coordinate (Q) at (-4,-1); 
\draw[red, ->] (O) -- (P) node[midway, above, yshift=0.3em] {$\av{H}_{\mathrm{irr}}$};

\draw[green!45!black, ->] (Q) -- (O) node[midway, below] {$\av{H}_{\mathrm{rev}}$};

\draw[green!45!black, ->] (Q) -- (P) node[midway, above] {$\av{H}$};

\fill[black] (O) circle (1.5pt);
\fill[black] (P) circle (1.5pt);
\fill[black] (Q) circle (1.5pt);

\end{tikzpicture}
    \caption{Pinning ellipsoid with semi-axes $\kappa_\mathrm{x}$, $\kappa_\mathrm{y}$, $\kappa_\mathrm{z}$. The reversible field $\av{H}_\mathrm{rev}$ points to the center, the applied field $\av{H}$ to a point on the surface, and $\av{H}_\mathrm{irr}$ connects the two.}
    \label{fig:ellipsoid}
\end{figure}
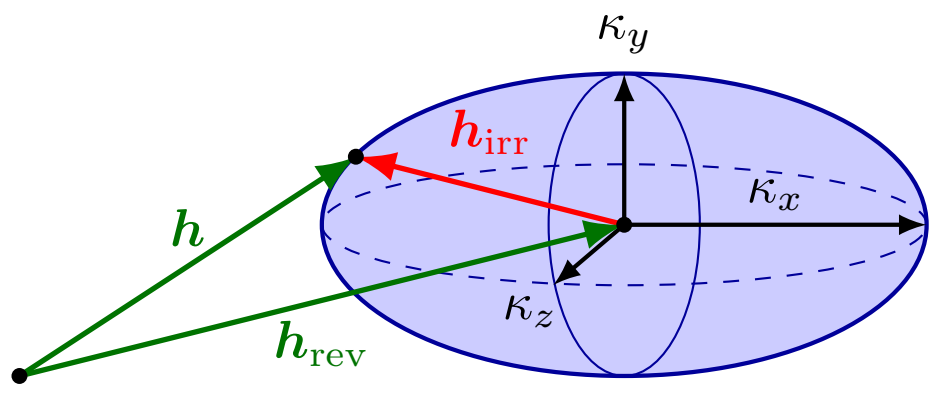

In the anisotropic extension, the scalar pinning strength $\kappa$ of each pinning particle is replaced by a diagonal matrix 
\begin{equation}
    \Material{\kappa} = \begin{bmatrix}
        \kappa_\mathrm{x} & 0 & 0\\
        0 & \kappa_\mathrm{y} & 0\\
        0 & 0 & \kappa_\mathrm{z}
        \end{bmatrix}~,
\end{equation}
which turns the spherical pinning particle of the isotropic model into the ellipsoid depicted in \autoref{fig:ellipsoid}~\cite{chen2024anisotropic}.

The reversible field stays unchanged as long as $\av{h}$ remains inside the pinning ellipsoid centered at $\av{h}_\mathrm{rev}$,
\begin{equation}
    \lVert\Material{\kappa}^{-1}(\av{h} - \av{h}_\mathrm{rev})\rVert \le 1.
\end{equation}
Otherwise, $\av{h}_\mathrm{rev}$ is updated such that (i) $\av{h}$ lies on the ellipsoid surface and (ii) $\av{h}_\mathrm{rev,new}$ lies on the line defined by $\av{h}-\av{h}_\mathrm{rev,old}$, as required by the VPM~\cite{jacques2018energy}. 

By just replacing $\kappa$ by $\Material{\kappa}$ in \eqref{eq:isotropic_vpm_H_rev}, as done in \cite{prigozhin2016hysteresis}, results in
\begin{equation}\label{eq:wrong_update}
    \av{h}_\mathrm{rev,new} = \av{h} - \Material{\kappa}\, \frac{\av{h}_\mathrm{irr}}{\lVert\av{h}_\mathrm{irr}\rVert}~,
\end{equation}
which scales the unit vector of $\av{h}_\mathrm{irr}$ by $\Material{\kappa}$.
This places $\av{h}_\mathrm{rev,new}$ on the ellipsoid surface, but for anisotropic $\Material{\kappa}$ the irreversible part $\av{h}-\av{h}_\mathrm{rev,new}$ is no longer parallel to $\av{h}_\mathrm{irr}$, violating the direction requirement (ii) of the VPM. 

The derivation for a geometrically consistent update formula starts with the condition that $\av{h}_\mathrm{rev,new}$ is on the ellipsoid
\begin{equation}\label{eq:ellipsoid condition}
  \lVert \Material{\kappa}^{-1}(\av{h} - \av{h}_\mathrm{rev,new})\rVert = 1~,
\end{equation}
to fulfill (i).
In order to fulfill (ii), $\av{h}_\mathrm{rev,old}$ needs to be moved towards $\av{h}$ in the direction of $\av{h}_\mathrm{irr}$
\begin{equation}\label{eq:ellipse update Hrev 1}
    \av{h}_\mathrm{rev,new} = \av{h}_\mathrm{rev,old} + s \av{h}_\mathrm{irr} = \av{h}_\mathrm{rev,old} + s (\av{h} - \av{h}_\mathrm{rev,old})~,
\end{equation}
with a length $s$.
Now by placing \eqref{eq:ellipse update Hrev 1} into \eqref{eq:ellipsoid condition} an expression for $s$ can be found by solving
\begin{gather}\label{eq:derivation_1}
      \lVert \Material{\kappa}^{-1}(\av{h} - \av{h}_\mathrm{rev,old} - s (\av{h} - \av{h}_\mathrm{rev,old})) \rVert = 1
\end{gather}
which gives
\begin{equation}
      s = 1 - \frac{1}{\lVert\Material{\kappa}^{-1}\av{h}_\mathrm{irr}\rVert}
\end{equation}

\begin{figure}[!hbt]
    \centering
    \includegraphics[width=\linewidth]{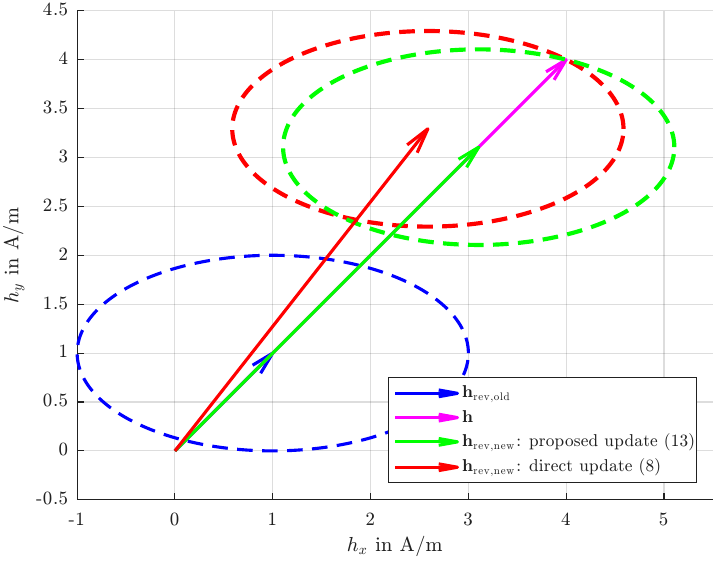}
    \caption{Update of $\av{H}_\mathrm{rev}$ in one time step: proposed formula \eqref{eq:update_formula} (in green) versus direct replacement of $\kappa$ by $\Material{\kappa}$ \eqref{eq:wrong_update} (in red).}
    \label{fig:visualization_ellipse}
\end{figure}

The solution for $s$ is put back into \eqref{eq:ellipse update Hrev 1} giving the proposed update formula
\begin{equation}\label{eq:update_formula}
        \av{h}_\mathrm{rev,new} = \av{h} - \frac{\av{h}_\mathrm{irr}}{\lVert\Material{\kappa}^{-1}\av{h}_\mathrm{irr}\rVert}~.
\end{equation}
For the complete derivation of \eqref{eq:update_formula} we refer to Appendix A.
\autoref{fig:visualization_ellipse} compares the two update rules geometrically and the violation of (ii) by \eqref{eq:wrong_update} can be observed.

For isotropic materials, $\Material{\kappa}=\kappa\mtr{I}$, both formulas reduce to \eqref{eq:isotropic_vpm_H_rev}. For anisotropic materials, only \eqref{eq:update_formula} satisfies both the boundary condition and the direction requirement (see \autoref{fig:visualization_ellipse}), which makes it the geometrically consistent generalization of \eqref{eq:isotropic_vpm_H_rev}.
\par
To illustrate this misalignment, the two update rules are compared in the case of unidirectional excitation. In this case, the VPM is exact, not an approximation of the energy-based model~\cite{jacques2018energy}.
The misalignment introduces an error that accumulates with every evaluation of \eqref{eq:wrong_update}.
\begin{figure}[!htb]
    \centering
    \includegraphics[width=0.8\linewidth]{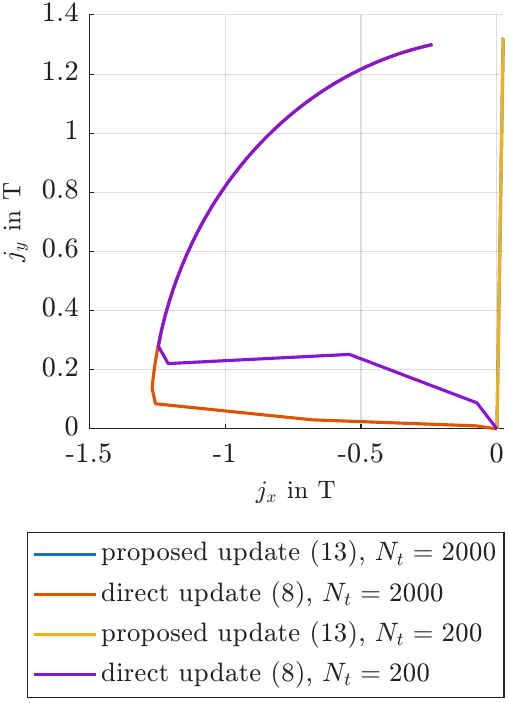}
    \caption{Locus of $\av{j}$ for a unidirectional excitation along the hard axis, tilted by $1^\circ$ toward the easy axis, with $\kappa_\mathrm{hard}/\kappa_\mathrm{easy}=0.1$: proposed versus direct update for different number of timesteps $N_t$.}
    \label{fig:comparison_lineplot_2}
\end{figure}
To demonstrate this error and to show that the proposed formula produces unidirectional results, we consider the following case: the excitation $\av{h}$ is unidirectional, directed along the hard axis (here the y axis), and tilted by $1^\circ$ towards the easy axis. A small misalignment of this kind is representative of numerical simulations, where it can occur even when the nominal demagnetizing field points along the hard axis.
For a strongly anisotropic ellipsoid, $\kappa_\mathrm{hard}/\kappa_\mathrm{easy}=0.1$, the resulting loci of $\av{j}$ for different numbers of timesteps $N_t$ are shown in \autoref{fig:comparison_lineplot_2}. Only the proposed formula yields a unidirectional locus. The locus from \eqref{eq:wrong_update} is not unidirectional, and the resulting J--H curve (\autoref{fig:comparison_j-h}) is therefore not physical.
\par
\begin{figure}[!htb]
    \centering
    \includegraphics[width=\linewidth]{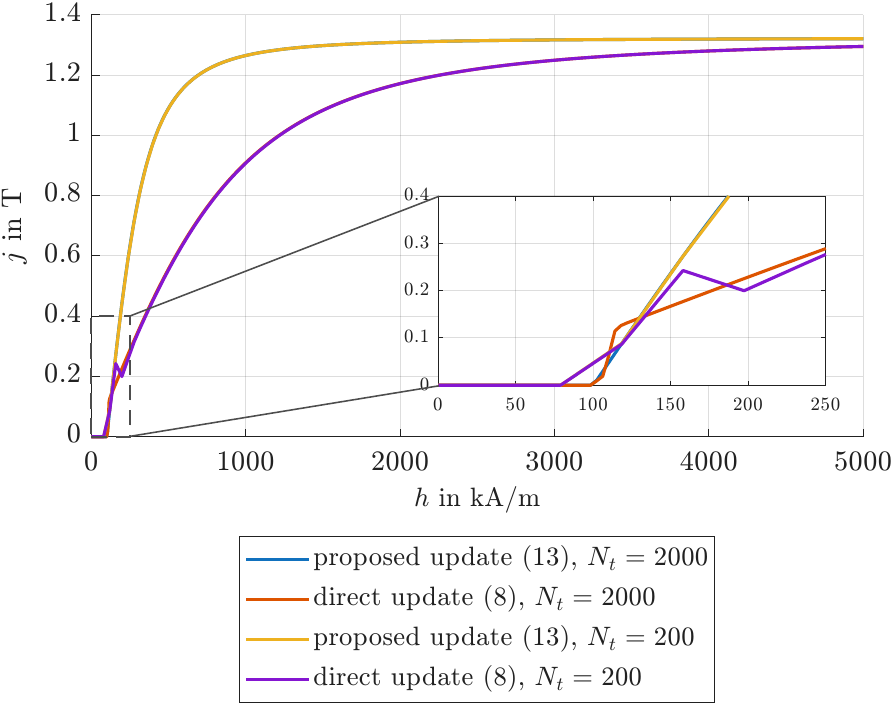}
    \caption{First-quadrant J--H curves for the excitation of \autoref{fig:comparison_lineplot_2}: proposed versus direct update for different number of timesteps $N_t$.}
    \label{fig:comparison_j-h}
\end{figure}
Furthermore, the direct update formula \eqref{eq:wrong_update} depends on the time discretization: both the loci of $\av{j}$ and the J--H curves change with $N_t$. It is therefore not rate independent. The deviations occur at low field values and vanish at higher field values. In contrast, the results of the proposed update formula are independent of $N_t$.
\par
\begin{figure}[!htb]
    \centering
    \includegraphics[width=\linewidth]{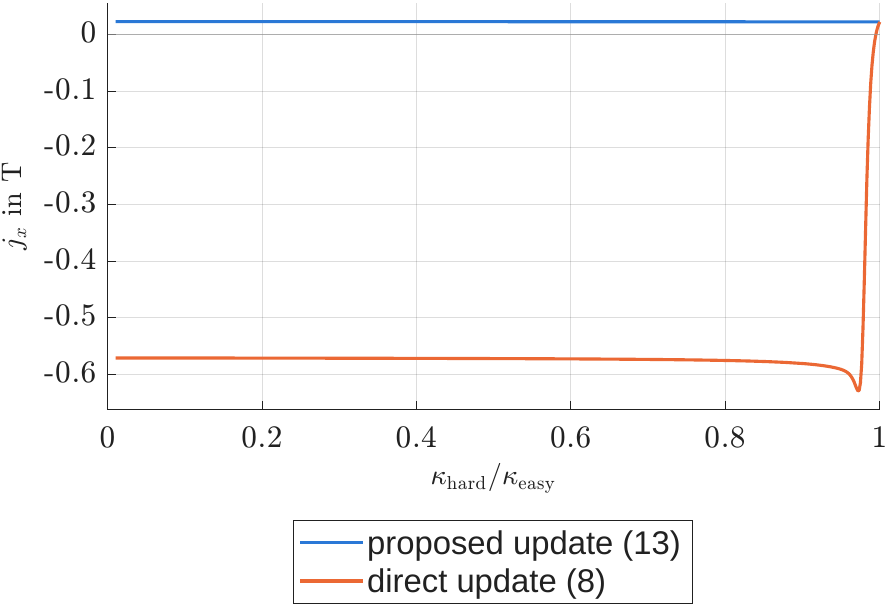}
    \caption{Comparison of $j_x$ versus $\kappa_\mathrm{hard}/\kappa_\mathrm{easy}$ for fixed $\av{h}$ tilted by $1^\circ$ toward the easy axis.}
    \label{fig:comparison_lineplot_3}
\end{figure}

The example above uses a strong anisotropy, so the influence of the ratio $\kappa_\mathrm{hard}/\kappa_\mathrm{easy}$ is examined next.
In \autoref{fig:comparison_lineplot_3}, the x component of $\av{j}$ is shown as a function of $\kappa_\mathrm{hard}/\kappa_\mathrm{easy}$ for a fixed $\av{h}$ and again a rotation of $1^\circ$ toward the easy axis. For the proposed update, the x component stays constant for different ellipsoids. For the update in \eqref{eq:wrong_update}, by contrast, $j_x$ is not constant with decreasing ratio, hence increasing anisotropy, and has the opposite sign.
The proposed update \eqref{eq:update_formula} thus remains consistent for all ratios $\kappa_\mathrm{hard}/\kappa_\mathrm{easy}$, whereas \eqref{eq:wrong_update} does not.

\subsection{Anisotropic Anhysteresis Function}
Aside from the anisotropic pinning forces, the anhysteresis function must be dependent on the direction of the applied field.
For this purpose, the modified Pacejka function~\cite{sauseng2025new}
\begin{equation}
  J_\mathrm{anh}(H) \;=\; J_\mathrm{s}\,\sin\!\bigl(\arctan(a H + b\arctan(cH))\bigr)~,
  \label{eq:magic}
\end{equation}
is used for both the easy and hard directions, each with its own set of parameters.
For field directions between the easy and hard axes, the anhysteretic parameters are interpolated
using a weight 
\begin{equation}
  w(t) \;=\; t^{\,n} \in [0,1],
  \label{eq:weight}
\end{equation}
which is a power law of the normalized angle $t$ between $\av{h}$ and the easy axis $\av{e}$
\begin{equation}
  \varphi \;=\; \arccos\bigl(\lvert \tp{\av{h}}\av{e}\rvert\bigr) \in [0,\tfrac{\pi}{2}],
  \quad t \;=\; \frac{\varphi}{\pi/2} \in [0,1]~.
  \label{eq:t-def}
\end{equation}
In doing so, each anhysteresis parameter $p \in \{J_\mathrm{s}, a, b, c\}$ is a convex combination of its easy- and hard-axis fit
\begin{equation}
  p(t) = p_\mathrm{easy}^{(1-w(t))} \cdot p_\mathrm{hard}^{w(t)}~.
  \label{eq:interp}
\end{equation}
The only additional fitting parameter is the exponent $n$ for every anhysteresis parameter, and the axis fits are unaffected because $w = 0$ and $w = 1$ on the easy and hard axes, respectively.

Instead of the modified Pacejka anhysteretic function~\cite{sauseng2025new}, other functions, for example the inverse tangent function, can be used. 

\section{Results}
The proposed model is fitted to pulsed-field measurements of an NdFeB magnet at different inclination angles $\phi$ of the applied field. The resulting J–H and B–H curves are then compared to the measured ones.
The model shows how well it can represent the measured behavior, not how well it can predict conditions it has not encountered before. This is because the same data are used for both fitting and comparison.
The angle between the applied field and the easy axis (inclination angle) was set to \SI{0}{\degree} (easy axis),
\SI{15}{\degree}, \SI{30}{\degree}, \SI{45}{\degree}, \SI{60}{\degree},
\SI{75}{\degree}, and \SI{90}{\degree} (hard axis). Prior to each measurement, the sample was magnetized with a \SI{7}{\tesla} pulse along the measurement direction using a Metis HyMPulse Pulsed Field Magnetometer.

The model parameters are fitted in two stages: first the anhysteresis, then the pinning. A single pinning cell is used because the measurements only include the major loop and no inner loops. In both stages the parameters $\boldsymbol\vartheta$ are determined by nonlinear least squares, which minimizes the deviation between the simulated and the measured polarization over all measured angles $\phi_i$:
\begin{equation}
  \min_{\boldsymbol\vartheta}\;\sum_{i=1}^{7}\sum_{m}
  \Bigl[J^{\mathrm{mod}}\bigl(H_{i,m};\boldsymbol\vartheta\bigr)-J^{\mathrm{meas}}_{i,m}\Bigr]^2 .
\end{equation}
In the first stage, $J^{\mathrm{mod}}=f\bigl(H;\mathbf p(\varphi_i)\bigr)$ is the anisotropic anhysteresis function and $J^{\mathrm{meas}}$ is the anhysteresis extracted from the measured loops as the mean of the ascending and descending branch fields at equal polarization. 
The fitted parameters $\boldsymbol\vartheta_1$ are the saturation polarization $J_s$ and the shape parameters $a$, $b$ and $c$ of the anhysteresis function, each on the easy and on the hard axis, together with one interpolation exponent $n_p$ per parameter that sets the transition between the two axes (12 parameters in total).
In the second stage, the anhysteresis is kept fixed. $J^{\mathrm{mod}}$ is the polarization component along the field simulated with the VPM along the measured major loop, and $J^{\mathrm{meas}}$ is the measured loop. The fitted parameters $\boldsymbol\vartheta_2$ are the semi-axes $\kappa_x$ and $\kappa_y$ of the pinning ellipsoid $\mathrm{diag}(\kappa_x,\kappa_y,\kappa_y)$.
\autoref{tab:anh_params} and \autoref{tab:fit_params_N1} list the fitted parameters.
\begin{table}[!htb]
\caption{Anisotropic Anhysteresis Parameters (Pacejka, Geometric Interpolation)}\label{tab:anh_params}
\begin{center}
\begin{tabular}{|c||c|c|c|c|}
\hline
&$J_\mathrm{s}$ in \si{\tesla}&$a$ in \si{\meter\per(\kilo\ampere)}&$b$&$c$ in \si{\meter\per(\kilo\ampere)}\\\hline
Easy&1.314&$2.070\cdot10^{-3}$&0.9349&$2.388\cdot10^{-2}$\\\hline
Hard&1.471&$2.246\cdot10^{-4}$&$7.160\cdot10^{-2}$&$9.949\cdot10^{-5}$\\\hline
$n$&0.2000&0.2081&1.741&2.937\\\hline
\end{tabular}
\end{center}
\end{table}

\begin{figure}[!hb]
    \centering
    \includegraphics[width=\linewidth]{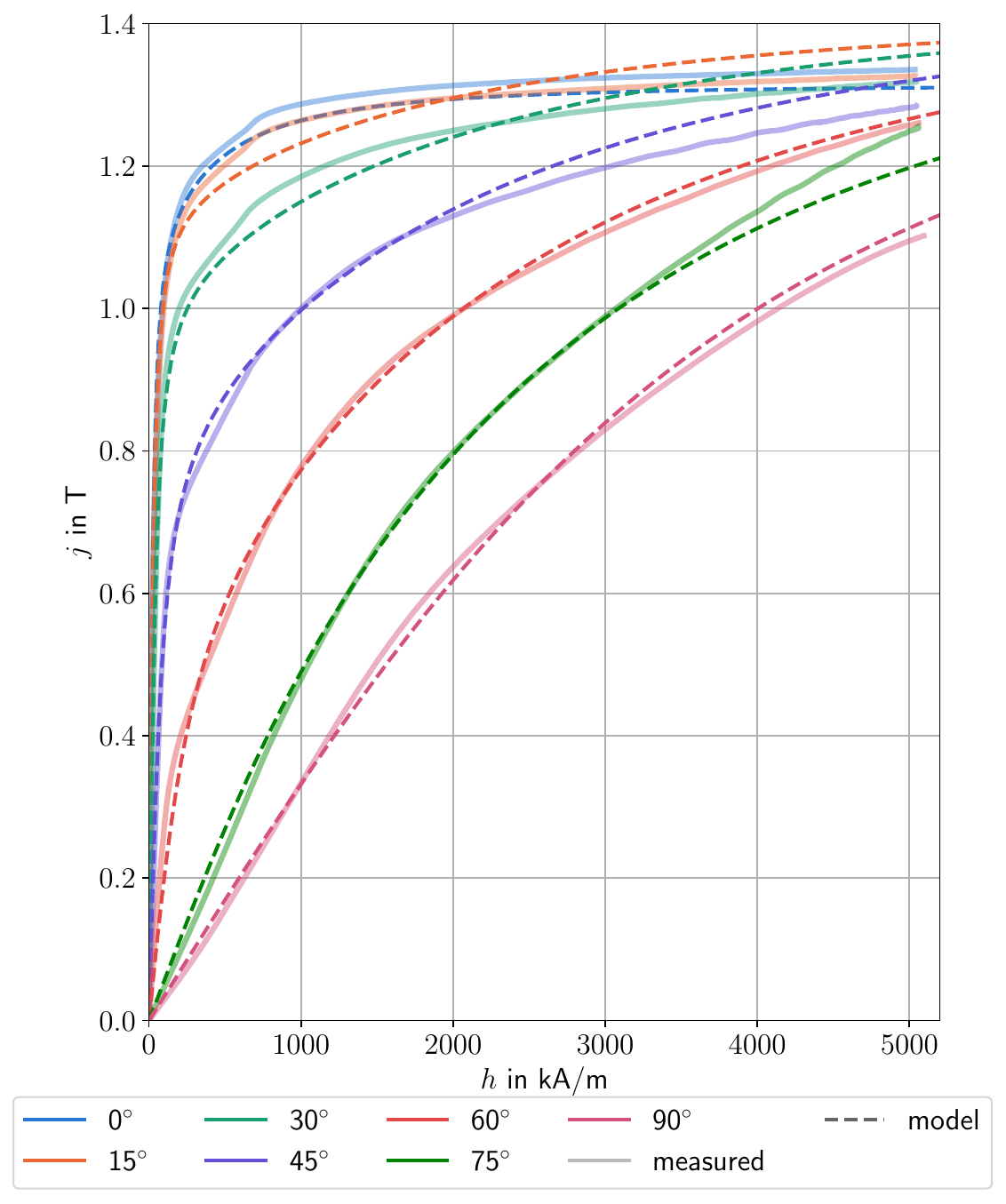}
    \caption{Measured and modeled anhysteresis functions for inclination angles between 0° and 90° of the appended field relative to the easy axis.}
    \label{fig:fitting_anhyst_result}
\end{figure}

\begin{table}[!htb]
\caption{Anisotropic Pinning Particle Parameters}\label{tab:fit_params_N1}
\begin{center}
\begin{tabular}{|c|c|}
\hline
Parameter&Value\\\hline
$\kappa_x$&\SI{989.7}{\kilo\ampere/\meter}\\\hline
$\kappa_y = \kappa_z$&\SI{903.3}{\kilo\ampere/\meter}\\\hline
\end{tabular}
\end{center}
\end{table}

\begin{figure}[!hb]
    \centering
    \includegraphics[width=\linewidth]{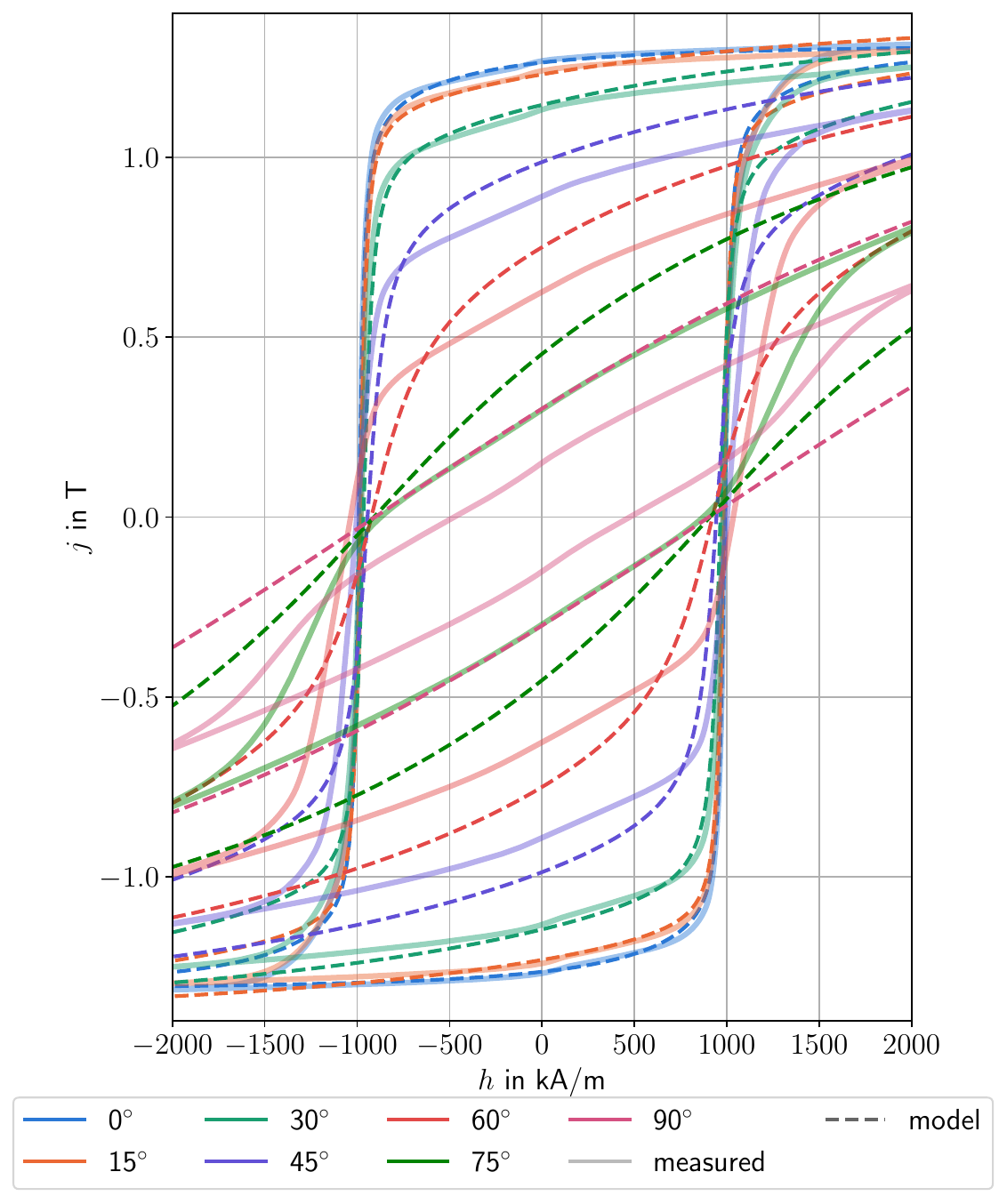}
    \caption{Measured and modeled J–H major loops for inclination angles between 0° and 90° of the appended field relative to the easy axis.}
    \label{fig:fitting_result}
\end{figure}

\begin{figure}[!htb]
    \centering
    \includegraphics[width=\linewidth]{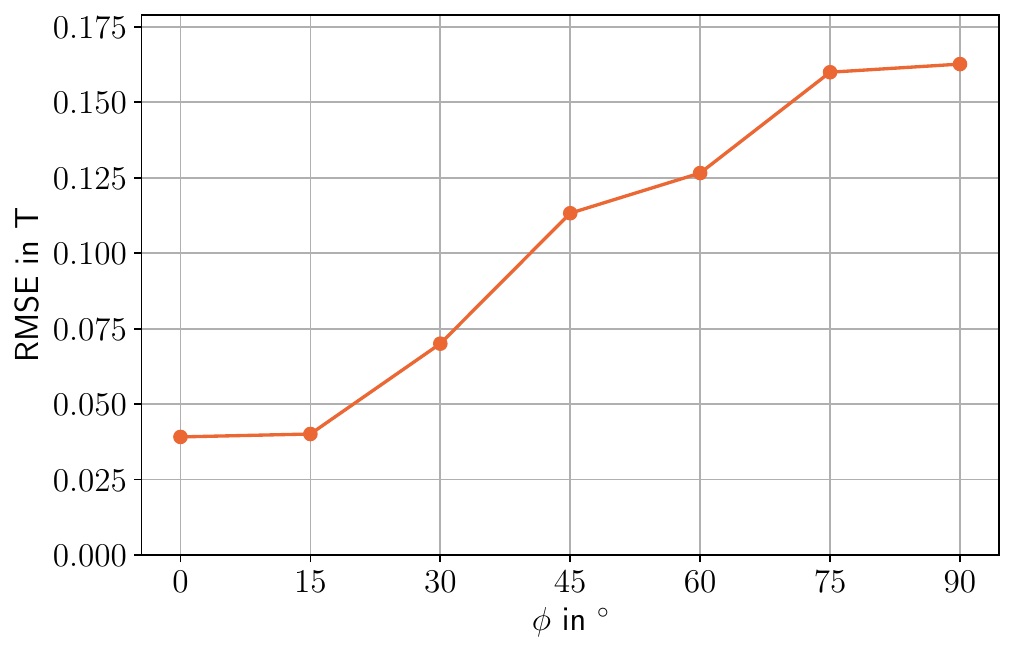}
    \caption{RMSE between the fitted model and the measurements.}
    \label{fig:rmse_over_angle}
\end{figure}

It can be observed that the ratio $\kappa_\mathrm{hard}/\kappa_\mathrm{easy} = 0.9127$
is near 1 (isotropic materal), because the coercivity is only decreasing for high inclination angles, as seen in \autoref{fig:fitting_result} where the measured and modeled J–H loops are shown. Please note that other materials the coercivity decreases more continous over the angles, as reported from the measurements of \cite{Campbell1982Alnico}.
For small inclination angles up to $30^\circ$, the model reproduces the measured curves well. For angles above $45^\circ$, however, it deviates from the measurements, which can be observed from the plot of the root mean square error (RMSE) of the fitted $\av{j}$ shown in \autoref{fig:rmse_over_angle}. 
A possible explanation is that the model does not represent the crystalline anisotropy explicitly: anisotropy enters only through the easy- and hard-axis anhysteretic fits and the pinning ellipsoid.

\begin{figure}[!htb]
    \centering
    \includegraphics[width=\linewidth]{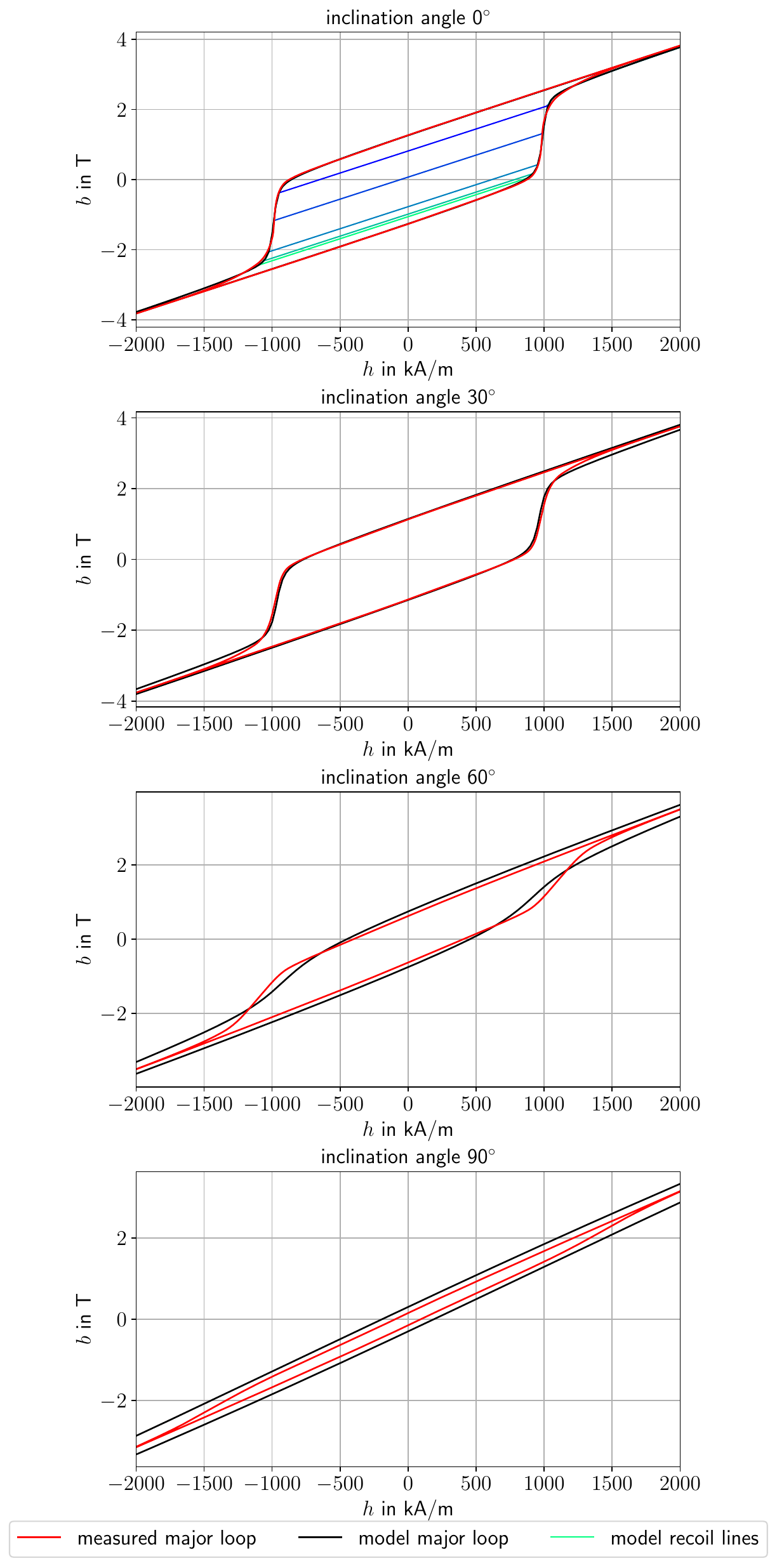}
    \caption{Measured and modeled B–H major loops with the modeled recoil lines for different inclination angles.}
    \label{fig:result_bh}
\end{figure}

These J–H deviations are less visible when the loops are evaluated in terms of the flux density $\av{b}$.
\autoref{fig:result_bh} shows the measured and modeled B–H major loops, as well as the recoil lines predicted by the model. Because a single-cell model is used, the inner loops are simply lines connecting the left and right branches of the major loop, essentially recoil lines.

\section{Conclusion}
A geometrically consistent, three-dimensional anisotropic extension of the VPM was presented, consisting of an ellipsoidal pinning region and an angle-dependent anhysteretic function. 
The model is by design three-dimensional, and since the isotropic VPM is already used in finite element formulations, this extension should also be usable in that context.

Recoil lines are obtained inherently from the single-cell model, whereas many existing PM models must switch between a major-loop description and a separate description of the recoil lines~\cite{Westerberg2024}.

Compared with measurements on an NdFeB magnet at inclination angles from \SI{0}{\degree} to \SI{90}{\degree}, the model reproduces the measured loops up to \SI{30}{\degree}.
Deviations remain at larger angles, which calls for further model extensions.

\section{Acknowledgement}
The work is supported by the joint DFG/FWF Collaborative Research Centre CREATOR (CRC – TRR361/F90) at TU Darmstadt, TU Graz and JKU Linz.

\section{APPENDIX A: Derivation of the proposed update formula}\label{appendix_a}
The complete step by step derivation of the proposed update formula, starting from \eqref{eq:derivation_1}, is the following.
\begin{equation}
    \lVert\Material{\kappa}^{-1}(\av{h} - \av{h}_\mathrm{rev,old} - s (\av{h} - \av{h}_\mathrm{rev,old}))\rVert = 1
\end{equation}
\begin{equation}
    \lVert\Material{\kappa}^{-1}(\av{h}_\mathrm{irr} - s \av{h}_\mathrm{irr})\rVert = 1
\end{equation}
\begin{equation}
    \lVert\Material{\kappa}^{-1}(1 - s)\av{h}_\mathrm{irr}\rVert = 1
\end{equation}
\begin{equation}
    \sqrt{\tp{(\Material{\kappa}^{-1}(1 - s)\av{h}_\mathrm{irr})}(\Material{\kappa}^{-1}(1 - s)\av{h}_\mathrm{irr})} = 1
\end{equation}
\begin{equation}
    (1-s)^2 \tp{(\Material{\kappa}^{-1}\av{h}_\mathrm{irr})}(\Material{\kappa}^{-1}\av{h}_\mathrm{irr}) = 1
\end{equation}
\begin{equation}
    (1-s)^2 = \frac{1}{\tp{(\Material{\kappa}^{-1}\av{h}_\mathrm{irr})}(\Material{\kappa}^{-1}\av{h}_\mathrm{irr})}
\end{equation}

\begin{equation}
    (1-s) = \sqrt{\frac{1}{\tp{(\Material{\kappa}^{-1}\av{h}_\mathrm{irr})}(\Material{\kappa}^{-1}\av{h}_\mathrm{irr})}}
\end{equation}
\begin{equation}
    (1-s) = \frac{1}{\lVert\Material{\kappa}^{-1}\av{h}_\mathrm{irr}\rVert}
\end{equation}
\begin{equation}
    s = 1 - \frac{1}{\lVert\Material{\kappa}^{-1}\av{h}_\mathrm{irr}\rVert}
\end{equation}

The solution for $s$ is put back into \eqref{eq:ellipse update Hrev 1}:
\begin{multline}\label{eq:corrected update formula}
        \av{h}_\mathrm{rev,new} = \av{h}_\mathrm{rev,old} + \left(1 - \frac{1}{||\Material{\kappa}^{-1}\av{h}_\mathrm{irr}||}\right) \av{h}_\mathrm{irr}\\ =
        \av{h}_\mathrm{rev,old} + \av{h}_\mathrm{irr} - \frac{\av{h}_\mathrm{irr}}{||\Material{\kappa}^{-1}\av{h}_\mathrm{irr}||} 
        = \av{h} - \frac{\av{h}_\mathrm{irr}}{||\Material{\kappa}^{-1}\av{h}_\mathrm{irr}||}
\end{multline}
\bibliographystyle{IEEEtran}
\bibliography{references}
\end{document}